\documentclass[aps,twocolumn,amsmath,amssymb]{revtex4-2}
\usepackage[dvipdfmx]{graphicx} 
\usepackage{bm}
\usepackage{braket}
\usepackage{hyperref}
\hypersetup{colorlinks=true, urlcolor=blue, linkcolor=blue, citecolor=blue}

\begin{document}

\title{Pairing properties of correlated three-leg ladders with strong interchain couplings \\ near 1/3 filling}

\author{Yushi Yamada, Tatsuya Kaneko, Masataka Kakoi, Ryota Ueda, and Kazuhiko Kuroki}

\affiliation{Department of Physics, The University of Osaka, Toyonaka, Osaka 560-0043, Japan}

\date{\today}

\begin{abstract}
We investigate the ground-state properties of correlated three-leg ladders near 1/3 filling. 
We apply the density-matrix renormalization group method to the three-leg $t$-$J$ ladder with strong interchain couplings and evaluate its pairing nature. 
When holes are doped into the spin-gapped state at 1/3 filling, we find that pair correlations develop with power-law decays while spin correlations decay exponentially. 
On the other hand, doping of electrons into the 1/3-filled state does not give rise to substantial pair correlations. 
We also discuss the hole-doped state in the three-leg Hubbard model to compare it with the pairing state in the $t$-$J$ model. 
Our numerical demonstrations provide insights into the electronic properties of trilayer nickelate superconductors. 
\end{abstract}

\maketitle

\section{Introduction}

The discovery of superconductivity (SC) in Ruddlesden-Popper-type nickelates has opened an intriguing research frontier for unconventional SC in strongly correlated systems. 
Following the first report of high-temperature SC in La$_3$Ni$_2$O$_7$ under pressure~\cite{HSun2023}, the emergence of SC has been discovered in various multilayer nickelates and their thin films~\cite{JHou2023,GWang2024,YZhang2024_NP,NWang2024,EKKo2025,GZhou2025}.
These findings stimulate numerous theoretical studies to elucidate its materials properties~\cite{ZLuo2023,QGYang2023,YShen2023,HOh2023,VChristiansson2023,YZhang2023,YFYang2023,FLechermann2023,YBLiu2023,ZLiao2023,XZQu2024,TKaneko2024,YCao2024,HSakakibara2024_327,YZhang2024_NC,GHeier2024,RJiang2024,CLu2024_1,WWu2024,LRhodes2024,JXWang2024,BGeisler2024,HLaBollita2024_PRB,MKakoi2024_DMRG,ZLuo2024,QGYang2024,CQChen2024,JChen2024,CLu2024_2,HYang2024,SRyee2024,JXZhang2024,YZhang2024_PRL,HSchlomer2024,HOh2025,YYZheng2025,JYYou2025,MOchi2025,KYJiang2025,TKaneko2025,SKamiyama2025,XZQu2025,KUshio_arXiv,SKamiyama_arXiv,HWatanabe_arXiv}. 
Although the transition temperature is lower than that of bilayer nickelates, SC in trilayer nickelates, such as La$_4$Ni$_3$O$_{10}$, have also attracted attention~\cite{HSakakibara2024_4310,QLi2024,JLi2024,YZhu2024,MKakoi2024_NMR,HNagata2024,MZhang2025,EZhang2025}. 
The electron configuration of Ni is considered as $d^{7.33\dots}$ in La$_4$Ni$_3$O$_{10}$, where the $d_{x^2-y^2}$ and $d_{3z^2-r^2}$ orbitals are mainly responsible for the electronic properties. 
In particular, the interlayer bonds of the $d_{3z^2-r^2}$ orbitals, whose network is nearly 1/3 filling~\cite{HSakakibara2024_4310}, are deeply involved in the characteristic electronic structures of trilayer nickelates. 
Given this background, understanding correlated many-body states in trilayer systems with strong interlayer couplings has become an important issue in relation to the study of nickelate superconductors. 

The three-leg ladder is a one-dimensional model that can mimic the pairing properties of trilayer systems. 
One advantage of using the three-leg ladder is that matrix-product-state-based methods can be applied effectively to investigating correlated many-body states. 
The single-orbital three-leg ladders have been studied, inspired by cuprates~\cite{EArrigoni1996_1,EArrigoni1996_2,TKimura1996,HLin1997,TRice1997,SWhite1998,TKimura1998,MKagan1999}. 
The phase diagram for the three-leg Hubbard ladder has already been drawn in the weak-coupling limit~\cite{EArrigoni1996_1,EArrigoni1996_2,HLin1997}. 
The carrier-doped state near half filling has been studied on the three-leg $t$-$J$ ladder with isotropic hopping, as an analog of the superconducting state in cuprates~\cite{TRice1997,SWhite1998}. 
On the other hand, there is a state in which both the charge and spin gaps are open (i.e., C0S0 phase) at 1/3 filling when interchain hopping is large~\cite{EArrigoni1996_1,EArrigoni1996_2,TKimura1998}. 
In the region slightly away from 1/3 filling, the weak-coupling theory predicts the existence of a state where one charge mode is gapless while the spin gap remains open (i.e., C1S0 phase), similar to the pairing state in the two-leg ladder near half filling~\cite{EDagotto1992,RNoack1994,MDolfi2015}. 
Since the doped state in this region potentially forms a pairing state associated with interchain bonds, understanding the doped state near 1/3 filling in the three-leg ladder is expected to provide meaningful insights into the electronic properties of trilayer nickelates. 

\begin{figure}[b]
\centering
\includegraphics[width=0.7\linewidth]{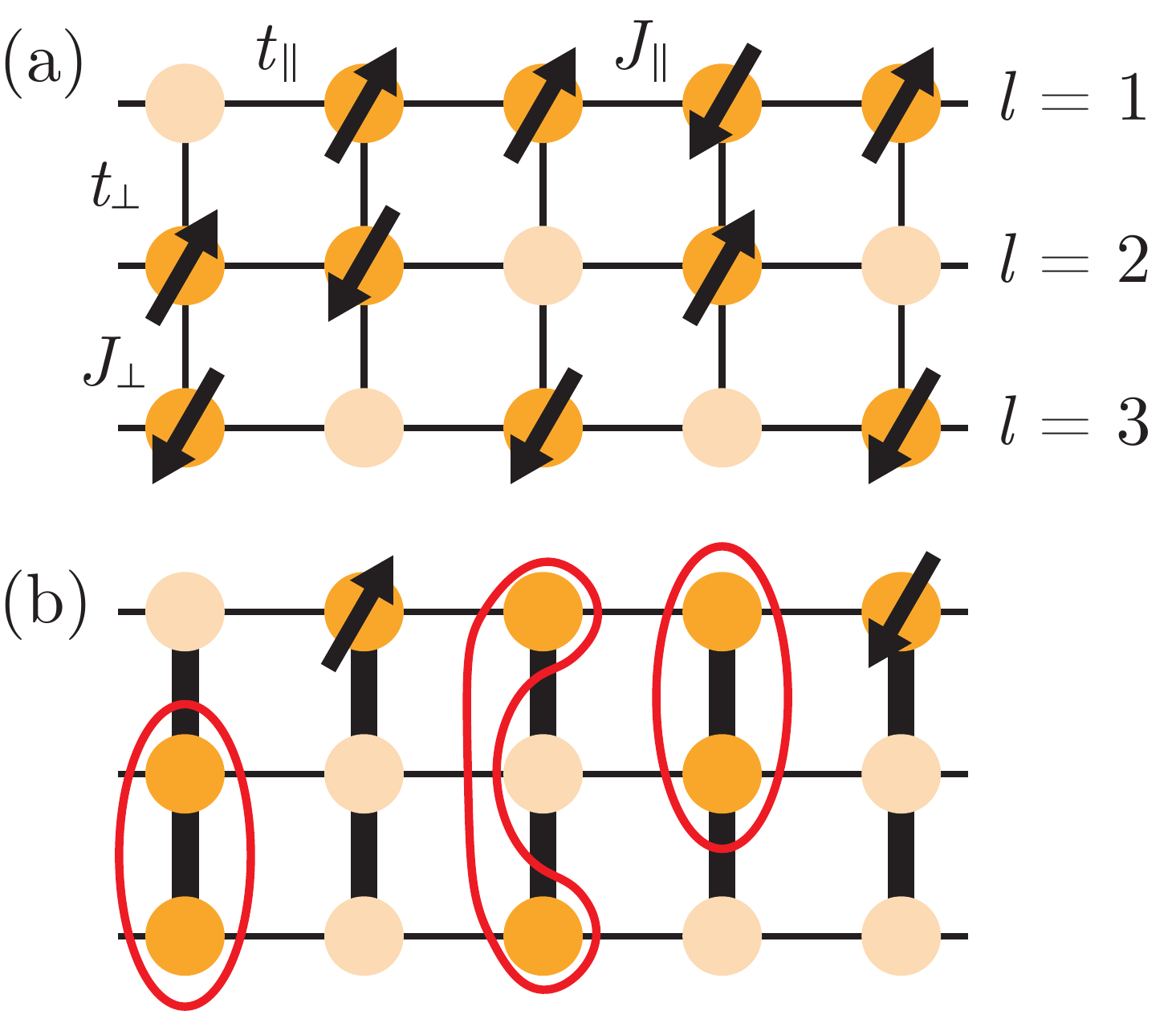}
\caption{(a) Three-leg $t$-$J$ ladder at 1/3 filling. (b) Hole-doped state near 1/3 filling with strong interchain coupling, where a pair of sites enclosed in a red continuous loop represents a spin singlet.} 
\label{fig1}
\end{figure}

In this paper, we investigate the pairing properties of correlated three-leg ladders near 1/3 filling. 
Employing the density-matrix renormalization group (DMRG) method, we calculate the pair correlation functions in the three-leg $t$-$J$ ladder. 
When holes are doped into the spin-gapped state at 1/3 filling, pair correlations develop with power-law decays, whereas spin correlations decay exponentially. 
Hence, the hole-doped state exhibits a signature favorable for SC. 
On the other hand, when electrons are doped into the 1/3-filled state, pair correlations do not exhibit substantial development. 
We also show changes in the ground-state properties when the strength of interchain coupling is varied in the hole-doped region. 
We compare it with the hole-doped state in the three-leg Hubbard model. 

The rest of this paper is organized as follows. 
In Sec.~\ref{sec:model_method}, we introduce the model and the numerical method employed in this study. 
In Sec.~\ref{sec:one_third_filling}, we present the spin correlation functions in the three-leg $t$-$J$ model at 1/3 filling. 
We show the pairing properties of the hole-doped and electron-doped states near 1/3 filling in Sec.~\ref{sec:hole_doping} and Sec.~\ref{sec:electron_doping}, respectively. 
In Sec.~\ref{sec:discussion}, we discuss the pairing tendency of the three-leg Hubbard model. 
A summary of our study is given in Sec.~\ref{sec:summary}.

\section{Model and Method} \label{sec:model_method}

We consider the three-leg $t$-$J$ ladder [Fig.~\ref{fig1}(a)], whose Hamiltonian is given by
\begin{align}
\hat{H} = 
& -t_{\parallel} \sum_{j} \sum_{l} \sum_{\sigma}
\left( \hat{\tilde{c}}^{\dagger}_{j, l, \sigma} \hat{\tilde{c}}_{j+1, l, \sigma} + \mathrm{H.c.} \right) 
\notag \\
& -t_{\perp} \sum_{j} \sum_{\langle l,l' \rangle} \sum_{\sigma}
\left( \hat{\tilde{c}}^{\dagger}_{j, l, \sigma} \hat{\tilde{c}}_{j, l', \sigma} + \mathrm{H.c.} \right) 
\notag \\
& +J_{\parallel} \sum_{j} \sum_{l} 
\left( \hat{\bm{S}}_{j, l}\cdot\hat{\bm{S}}_{j+1, l} - \frac{1}{4}\hat{n}_{j, l}\hat{n}_{j+1, l} \right) 
\notag \\
& +J_{\perp} \sum_{j} \sum_{\langle l,l' \rangle} 
\left( \hat{\bm{S}}_{j, l}\cdot\hat{\bm{S}}_{j, l'} - \frac{1}{4}\hat{n}_{j, l}\hat{n}_{j, l'} \right). 
\label{eq:H_t-J}
\end{align}
$\hat{c}^{\dagger}_{j, l, \sigma}$ ($\hat{c}_{j, l, \sigma}$) is the creation (annihilation) operator of an electron with spin $\sigma$~(=$\uparrow,\downarrow$) at site $j$ on chain $l$~$(=1,2,3)$. 
The operator $\hat{\tilde{c}}^{\dagger}_{j, l, \sigma} = \hat{c}^{\dagger}_{j, l, \sigma} ( 1 - \hat{n}_{j, l, \bar{\sigma}} )$ prohibits the creation of an doubly occupied site, where $\hat{n}_{j, l, \sigma} = \hat{c}^{\dagger}_{j, l, \sigma} \hat{c}_{j, l, \sigma}$ and $\bar{\sigma}$ is the opposite spin of $\sigma$. 
$\hat{n}_{j, l} = \hat{n}_{j, l, \uparrow} + \hat{n}_{j, l, \downarrow}$ counts the number of electrons at $(j,l)$, and $\hat{\bm{S}}_{j, l}$ is the spin operator. 
$t_{\parallel}$ and $t_{\perp}$ are the intrachain and interchain hopping parameters, respectively. 
$J_{\parallel}$ and $J_{\perp}$ are the coupling constants of the intrachain and interchain spin interactions, respectively. 
We denote the number of electrons as $N$ and the length of the chain as $L_x$ (where the lattice constant is set to 1). 
We define the electron density as $n = N/(3L_x)$, where $3L_x$ is the number of lattice sites and $n=1$ ($N=3L_x$) corresponds to half filling. 

To obtain the ground state of the model, we employ the DMRG method~\cite{SWhite1992,SWhite1993,USchollwock2011}. 
We use the three-leg ladder with open boundary conditions. 
In our calculation, we set $t_{\parallel}$ as the unit of energy. 
Unless otherwise specified, we present the results for $L_x=80$, $J_{\parallel}=t_{\parallel}/3$, and $J_{\perp}/J_{\parallel}=(t_{\perp}/t_{\parallel})^2$. 
The truncation errors were less than $1 \times 10^{-6}$ in most calculations, and the largest truncation error was $\sim 5 \times 10^{-6}$ (when $t_{\perp}/t_{\parallel}=1.5$ and $n=2/3-1/60$ in Fig.~\ref{fig5}), where the bond dimension was up to $m=5000$.

\section{Results}

\subsection{1/3 filling} \label{sec:one_third_filling}

First, we consider the ground state at 1/3 filling ($n=2/3$), where electrons occupy two of the three sites at $j$ on average. 
When $t_{\parallel}=J_{\parallel}=0$, the eigenstates of each unit cell can be determined by the three-site model exactly. 
In the three-site system, the lowest-energy state (when $J_{\perp}>0$) is given by 
\begin{align}
&\frac{u}{2} \left( \ket{\uparrow,\downarrow,0} - \ket{\downarrow,\uparrow,0} \right)
+\frac{u}{2} \left( \ket{0,\uparrow,\downarrow} - \ket{0,\downarrow,\uparrow} \right)
\notag \\
&+\frac{v}{\sqrt{2}} \left( \ket{\uparrow,0,\downarrow} - \ket{\downarrow,0,\uparrow} \right), 
\label{eq:three_site}
\end{align}
where $u^2+v^2=1$. 
In the $t$-$J$ model, $u^2= (1/2) (1 + J_{\perp} / \sqrt{J_{\perp}^2+8t_{\perp}^2} )$, see Appendix~\ref{appendix_A} for details.
The interchain spin singlets in the three-site state are schematically shown in Fig.~\ref{fig1}(b). 
Due to the magnetic interaction $J_{\perp}$, two adjacent sites form a spin singlet, which is similar to the two-leg spin ladder. 
Additionally, since $t_{\perp}$ allows single-particle hopping, the lowest-energy state of Eq.~\eqref{eq:three_site} also contains $\ket{\uparrow,0,\downarrow}$ and $\ket{\downarrow,0,\uparrow}$. 
Although the above three-site state is no longer the exact eigenstates when $t_{\parallel}$ ($J_{\parallel}$) is introduced, its characteristics may remain when $t_{\perp} > t_{\parallel}$ ($J_{\perp} > J_{\parallel}$), where the spin-singlet nature in each unit cell likely suppress the development of spin correlations along the chain direction. 

\begin{figure}[b]
\centering
\includegraphics[width=1.0\linewidth]{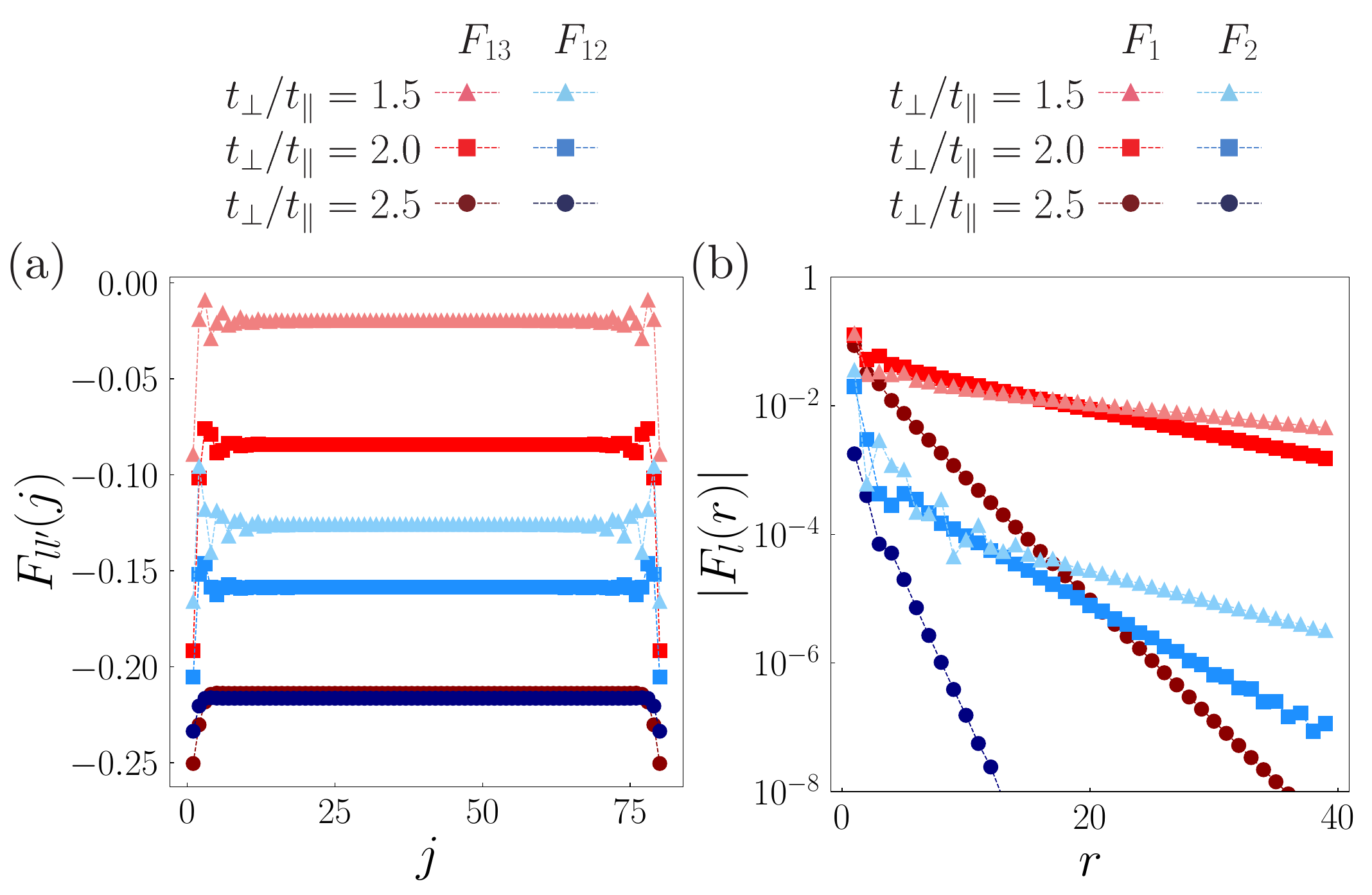}
\caption{(a) Interchain spin correlation function $F_{ll'}(j)$ ($=\braket{ \hat{\bm{S}}_{j, l} \cdot \hat{\bm{S}}_{j, l'} }$) and (b) intrachain spin correlation function $F_{l}(r)$ in the three-leg $t$-$J$ ladder at 1/3 filling ($n=2/3$).}
\label{fig2}
\end{figure}

To examine the above properties in the three-leg $t$-$J$ ladder, we compute the spin correlation functions at 1/3 filling ($n=2/3$). 
Figure~\ref{fig2}(a) shows the spin correlations along the rung ($\perp$) direction defined as 
\begin{align}
F_{ll'}(j) = \Braket{ \hat{\bm{S}}_{j, l} \cdot \hat{\bm{S}}_{j, l'} }
\end{align}
at $j$ with $l \ne l'$. 
The interchain spin correlations $F_{12}(j)$ and $F_{13}(j)$ negatively increase with $t_{\perp}/t_{\parallel}$. 
When $t_{\perp}/t_{\parallel}=2.5$, the spin correlations get closer to the values for the three-site state of Eq.~\eqref{eq:three_site}, i.e., $\braket{\hat{\bm{S}}_{1}\cdot\hat{\bm{S}}_{2}} = -3u^2/8 \approx -0.24$ and $\braket{\hat{\bm{S}}_{1}\cdot\hat{\bm{S}}_{3}} = -3v^2/4 \approx -0.27$. 
This suggests that the characteristics of the three-site state can appear in the large $t_{\perp}$ regime even with $t_{\parallel} \ne 0$. 

Figure~\ref{fig2}(b) plots the spin correlations along the chain ($\parallel$) direction defined as 
\begin{align}
F_{l}(r) = \Braket{ \hat{\bm{S}}_{j_0, l} \cdot \hat{\bm{S}}_{j_0+r, l} }, 
\end{align}
where $j_0$ is the reference site. 
We set $j_0=L_x/4+1$ to minimize the effects of open boundaries. 
The spin correlation functions in both the outer ($l=1$) and inner ($l=2$) chains exhibit exponential-like decays. 
As $t_{\perp}/t_{\parallel}$ increases, the spin correlations in $F_{1}(r)$ and $F_{2}(r)$ decay more steeply. 
This tendency corresponds to the enhancement of the interchain spin correlations in $F_{12}(j)$ and $F_{13}(j)$, indicating that the singlet-like configuration in each unit cell [Eq.~\eqref{eq:three_site}], which suppresses the spin correlations along the chain direction, is more strongly established as $t_{\perp}/t_{\parallel}$ increases. 
When the intrachain spin correlations $F_{1}(r)$ and $F_{2}(r)$ are short-ranged, correlations between interchain spin-singlet pairs potentially develop in the regions slightly away from 1/3 filling.

\subsection{Hole doping} \label{sec:hole_doping}

To investigate the pairing properties of doped states near 1/3 filling, we compute the pair correlation function 
\begin{align}
P_{ll'}(r) &= \Braket{ \hat{\Delta}_{j_0, ll'}^{\dagger} \hat{\Delta}_{j_0+r, ll'} } , 
\end{align}
where $\hat{\Delta}_{j, ll'} = (\hat{c}_{j, l, \uparrow}\hat{c}_{j, l', \downarrow} - \hat{c}_{j, l, \downarrow}\hat{c}_{j, l', \uparrow}) / \sqrt{2}$ is the operator for a spin-singlet pair formed by electrons on chains $l$ and $l'$. 
We set the reference site to $j_0=L_x/4+1$ as in $F_{l}(r)$. 
$P_{12}(r)$ represents correlations of pairs formed between the outer ($l=1$) and inner ($l=2$) chains, while $P_{13}(r)$ reflects those between the outer ($l=1$) and outer ($l=3$) chains. 
Here, we use $t_{\perp}/t_{\parallel}=2.5$ to investigate the case where the intrachain spin correlations are short-ranged due to the interchain spin singlets. 

\begin{figure}[t]
\centering
\includegraphics[width=1.0\linewidth]{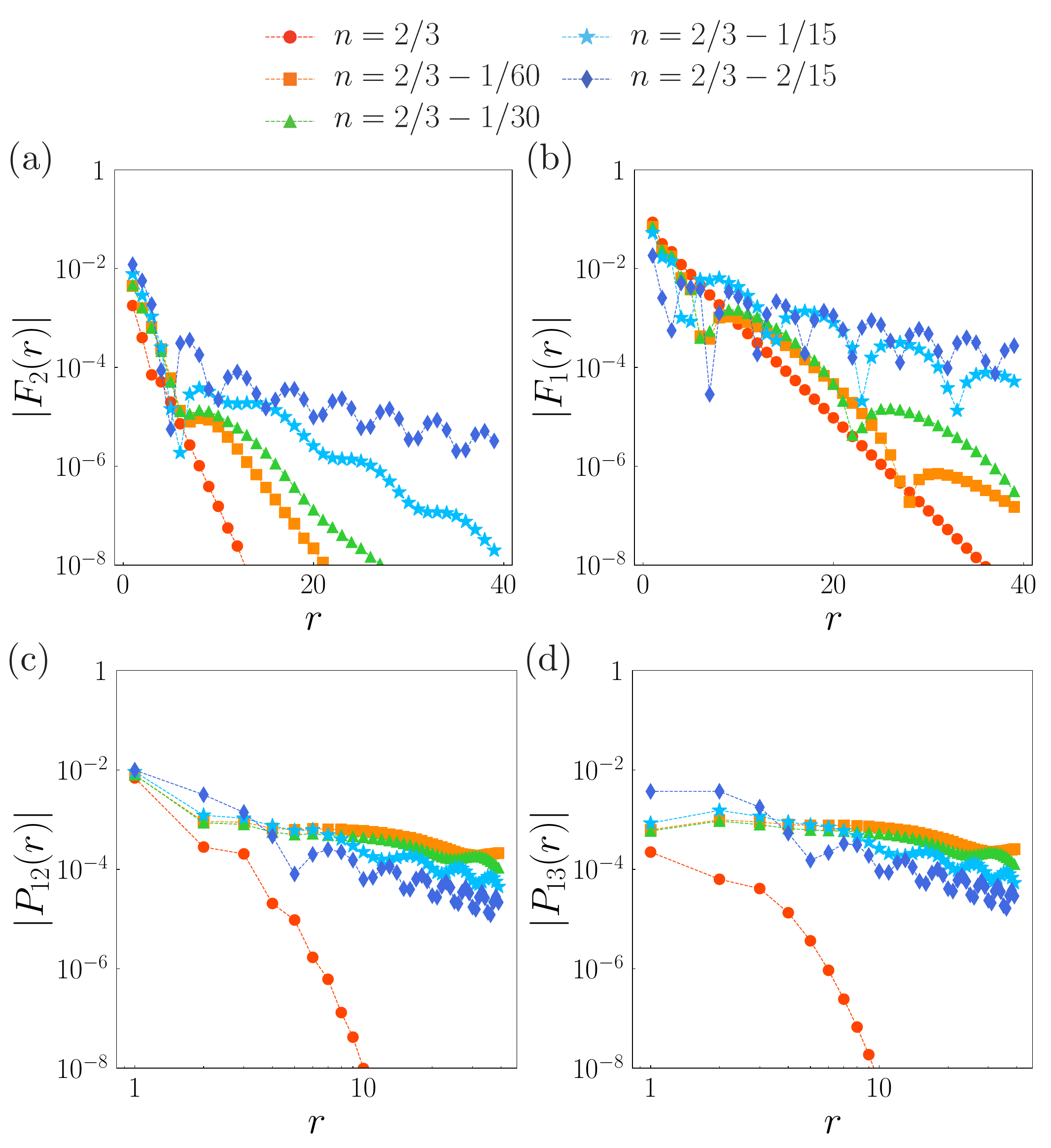}
\caption{Correlation functions for $t_{\perp}/t_{\parallel}=2.5$ in the hole-doped states near 1/3 filling. 
Top: Spin correlation functions (a) $F_{2}(r)$ for the inner chain and (b) $F_{1}(r)$ for the outer chain. 
Bottom: Pair correlation functions (c) $P_{12}(r)$ for pairs between the outer and inner chains and (d) $P_{13}(r)$ for pairs between the two outer chains.} 
\label{fig3}
\end{figure}

Figure~\ref{fig3} presents the spin and pair correlation functions in the hole-doped states away from 1/3 filling. 
Figures~\ref{fig3}(a) and \ref{fig3}(b) shows the spin correlation functions of the inner ($l=2$) and outer ($l=1$) chains in a semi-logarithmic plot. 
While the spin correlations gradually increase with hole doping, the exponential-like decay is maintained, suggesting that the spin-singlet nature remains even with hole doping. 
In contrast to the spin correlations, the pair correlations in $P_{12}(r)$ and $P_{13}(r)$ rise upon hole doping, as shown in Figs.~\ref{fig3}(c) and \ref{fig3}(d). 
The pair correlations decay linearly in a log-log plot, indicating a power-law decay. 
$P_{12}(r)$ exhibits a similar behavior with $P_{13}(r)$, corresponding that both the interchain spin correlations $F_{12}(j) = \braket{ \hat{\bm{S}}_{j, 1} \cdot \hat{\bm{S}}_{j, 2} }$ and $F_{13}(j) = \braket{ \hat{\bm{S}}_{j, 1} \cdot \hat{\bm{S}}_{j, 3} }$ are equally developed when $t_{\perp}/t_{\parallel}=2.5$ at 1/3 filling [Fig.~\ref{fig2}(a)]. 
The pair correlations develop over longer distances than the spin correlations in the low-doping region, whereas the pair correlations are gradually suppressed as the spin correlations increase with doping. 

To quantify the decaying tendency, we fit the pair correlation functions using $P_{ll'}(r) = A_0r^{-K_0} + A_1r^{-K_1}\cos(qr+\phi)$~\cite{XLu2023,YShen2023PRB,TKaneko2024JPSJ}. 
As seen in Figs.~\ref{fig3}(c) and \ref{fig3}(d), the wavenumber $q$ in the second term is proportional to the doping concentration $\delta_h=2/3-n$. 
For the fitting, we used the data of $P_{ll'}(r)$ at $10 \le r < L_x/2$. 
Because the amplitude $A_0$ is larger than $A_1$, we regard $K_0$ as the decay exponent $K_{\rm SC}$ of the pair correlation. 
For both $P_{12}$ and $P_{13}$, the estimated decay exponents at $\delta_h=1/30, 1/15$, and $2/15$ are $K_{\rm SC} =$ 0.9, 1.2, and 1.3, respectively. 
While $K_{\rm SC}$ increases with $\delta_h$, the values of $K_{\rm SC}$ is comparable to those in the two-leg Hubbard and $t$-$J$ ladders~\cite{MDolfi2015,XLu2023,YShen2023PRB}. 
Therefore, hole doping into the 1/3-filled state can induce a signature favorable for SC.

\subsection{Electron doping} \label{sec:electron_doping}

\begin{figure}[t]
\centering
\includegraphics[width=1.0\linewidth]{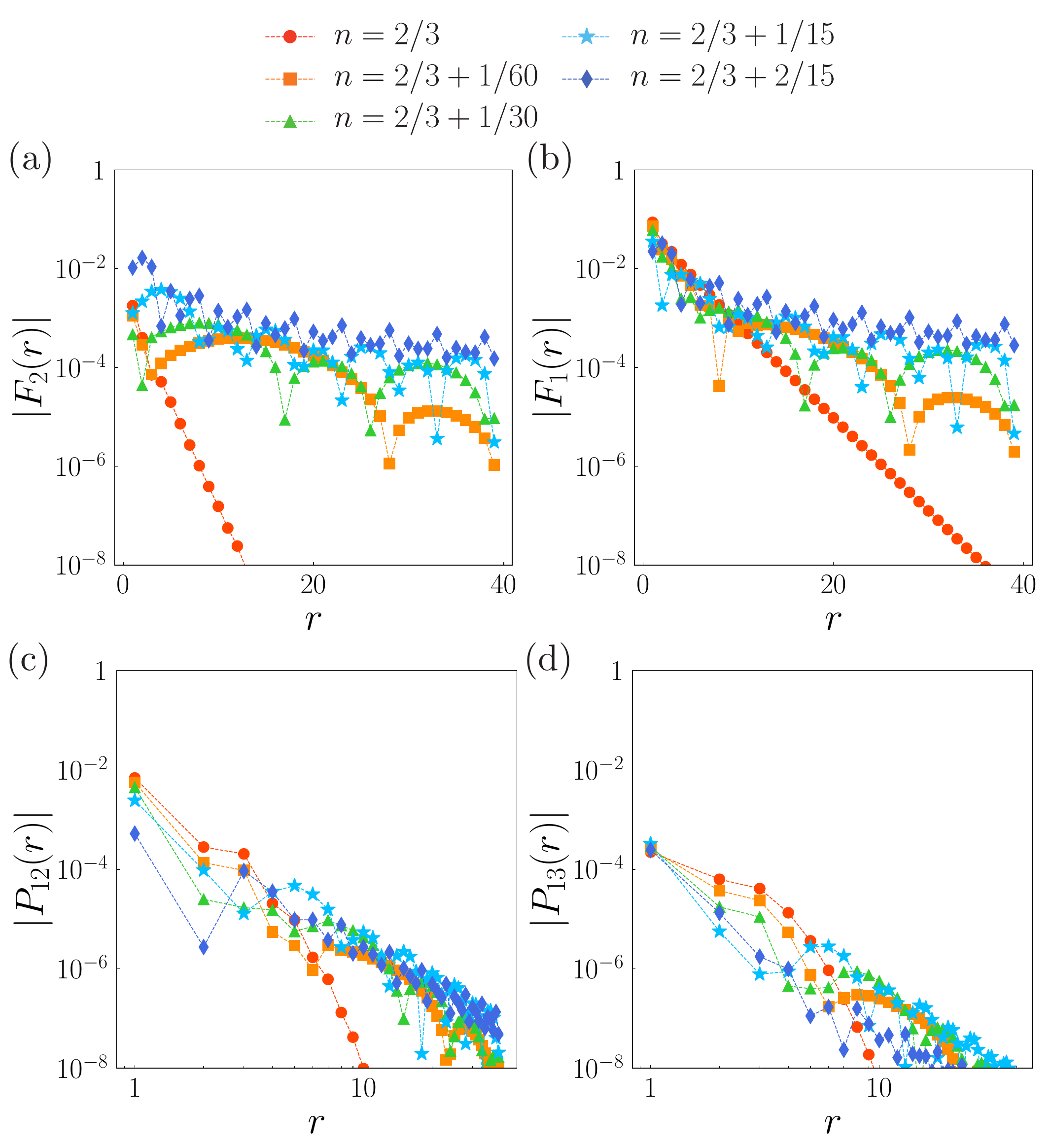}
\caption{Correlation functions for $t_{\perp}/t_{\parallel}=2.5$ in the electron-doped states near 1/3 filling. 
Top: Spin correlation functions (a) $F_{2}(r)$ for the inner chain and (b) $F_{1}(r)$ for the outer chain. 
Bottom: Pair correlation functions (c) $P_{12}(r)$ for pairs between the outer and inner chains and (d) $P_{13}(r)$ for pairs between the two outer chains.} 
\label{fig4}
\end{figure}

Next, we consider the case when electrons are doped into the 1/3-filled state. 
Figure~\ref{fig4} presents the results for the electron-doped states near 1/3 filling.
The increase in the spin correlations with electron doping [Figs.~\ref{fig4}(a) and \ref{fig4}(b)] is larger than that with hole doping. 
For large doping concentrations, the decay is no longer exponential-like. 
Compared to the case of hole doping [Fig.~\ref{fig3}(a)], the spin correlations in the inner chain [Fig.~\ref{fig4}(a)] are similarly developed as in the outer chains [Fig.~\ref{fig4}(b)]. 
$u^2/2$ and $v^2$ in the three-site state of Eq.~\eqref{eq:three_site} become close values when $t_{\perp}/t_{\parallel}=2.5$, suggesting that a hole is present on each of the three sites with a similar probability at 1/3 filling. 
The similar behavior of spin correlations in $F_{1}(r)$ and $F_{2}(r)$ with electron doping is due to the ability of doped electrons to reside equally in each chain. 

As shown in Figs.~\ref{fig4}(c) and \ref{fig4}(d), the pair correlations are not substantially enhanced by electron doping. 
In contrast to the case of hole doping, the spin correlations are much larger than the pair correlation, indicating that the long-ranged pairing state is unlikely to emerge with electron doping. 
In the case where the on-site repulsion $U$ is active and double occupancy is strongly disfavored (or prohibited in the $t$-$J$ model), the generation of the pair correlation requires the presence of unit cells with two or three holes (one or zero electrons). 
Hole doping creates such situations, whereas electron doping increases the number of unit cells without a hole (or three singly occupied sites). 
This results in less development of pair correlations in the case of electron doping. 
Hence, introducing carriers into the spin-gapped state at 1/3 filling does not necessarily lead to the development of the pair correlations. 
The difference between electron-doping and hole-doping near 1/3 filling is qualitatively consistent with the analytical assessment in the $t_{\perp}, J_{\perp} \gg t_{\parallel}, J_{\parallel}$ limit~\cite{MKagan1999}. 
The asymmetric pairing nature across the 1/3-filling line in the $t$-$J$ model differs from the phase diagram predicted in the weak-coupling ($U \rightarrow 0$) limit, which exhibits the C1S0 phases in both the electron-doped and hole-doped sides~\cite{EArrigoni1996_1,EArrigoni1996_2,HLin1997}.

\section{Discussion} \label{sec:discussion}

The $t$-$J$ model is an effective model of the Hubbard model in the strong-coupling limit. 
In this section, we discuss the pairing properties in the three-leg Hubbard model. 
The Hamiltonian of the three-leg Hubbard ladder reads 
\begin{align}
\hat{H} = 
& -t_{\parallel} \sum_{j} \sum_{l} \sum_{\sigma}
\left( \hat{c}^{\dagger}_{j, l, \sigma} \hat{c}_{j+1, l, \sigma} + \mathrm{H.c.} \right) 
\notag \\
& -t_{\perp} \sum_{j} \sum_{\langle l,l' \rangle} \sum_{\sigma} 
\left( \hat{c}^{\dagger}_{j, l, \sigma} \hat{c}_{j, l', \sigma} + \mathrm{H.c.} \right) 
\notag \\
& +U \sum_{j} \sum_{l} \hat{n}_{j, l, \uparrow} \hat{n}_{j, l, \downarrow}, 
\end{align}
where $U$ is the on-site Coulomb repulsion. 

\begin{figure}[t]
\centering
\includegraphics[width=1.0\linewidth]{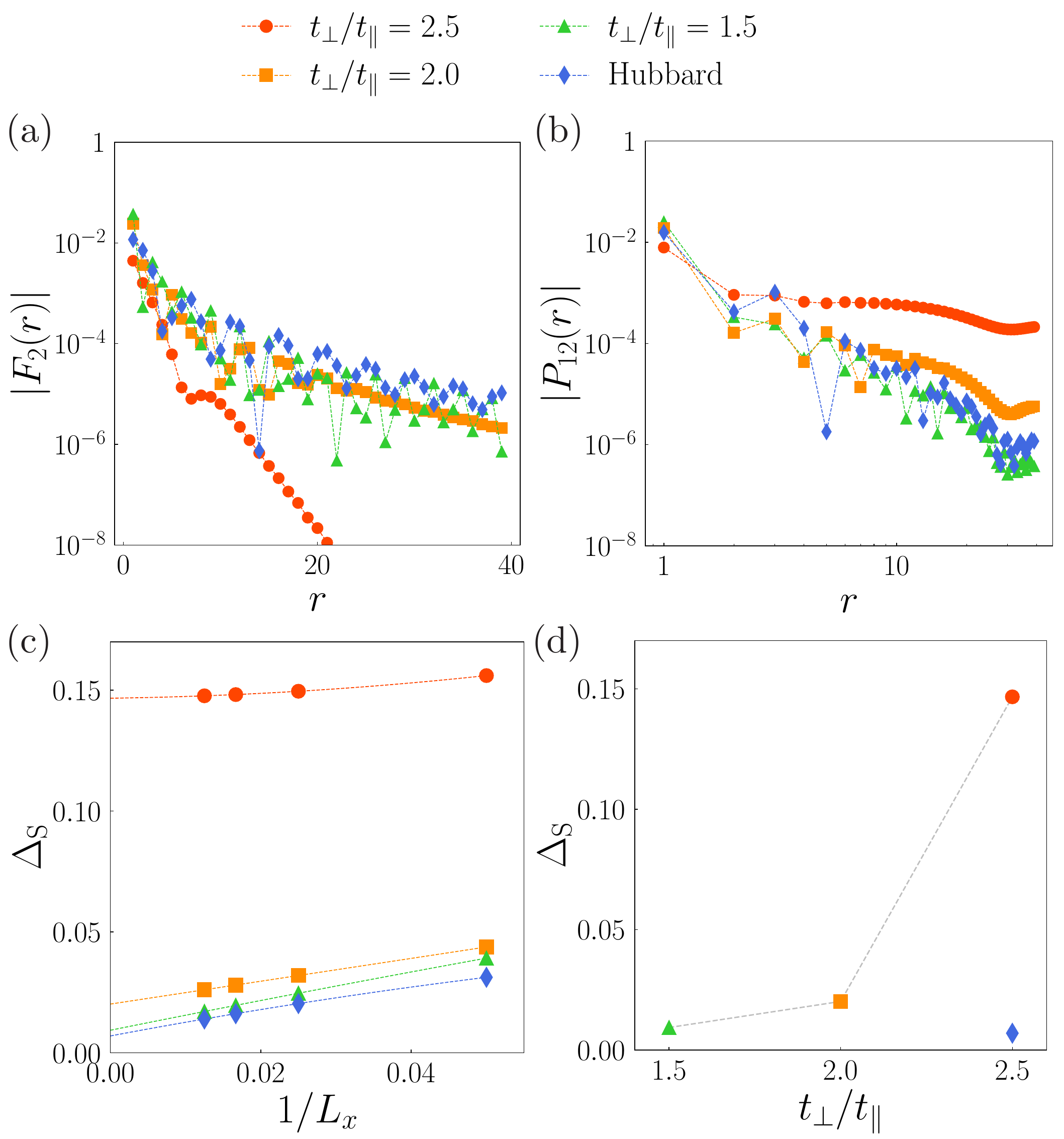}
\caption{(a) Spin and (b) pair correlation functions in the $t$-$J$ model for various $t_{\perp}/t_{\parallel}$ values and the Hubbard model for $t_{\perp}/t_{\parallel}=2.5$ and $U/t_{\parallel}=10$ at $n=2/3-1/60$. 
(c) Extrapolation of the spin gap $\Delta_{\rm S}$ at 1/3 filling ($n=2/3$) and (d) $t_{\perp}/t_{\parallel}$-dependence of $\Delta_{\rm S}$ in the thermodynamics limit ($1/L_x\rightarrow 0$).}
\label{fig5}
\end{figure}

Here, we compare the results of the Hubbard model for $t_{\perp}/t_{\parallel}=2.5$ and $U/t_{\parallel}=10$ with the results of the $t$-$J$ model. 
Figures~\ref{fig5}(a) and \ref{fig5}(b) show the spin and pair correlation functions in the hole-doped states (at $\delta_h = 2/3-n = 1/60$). 
In the $t$-$J$ model, the spin correlations increase as $t_{\perp}/t_{\parallel}$ decreases [Fig.~\ref{fig5}(a)]. 
The pair correlations in the $t$-$J$ model [Fig.~\ref{fig5}(b)] are suppressed as the spin correlations increase with the decrease in $t_{\perp}/t_{\parallel}$. 
Hence, the larger $t_{\perp}/t_{\parallel}$ ($J_{\perp}/J_{\parallel}$) is suitable for developing the pair correlation. 
Compared with the results in the $t$-$J$ model, the pair correlation in the Hubbard model is not well developed at $t_{\perp}/t_{\parallel}=2.5$. 
Both the spin and pair correlation functions in the Hubbard model at $t_{\perp}/t_{\parallel}=2.5$ are comparable to those in the $t$-$J$ model at $t_{\perp}/t_{\parallel}=1.5$. 

In Figs.~\ref{fig5}(c) and \ref{fig5}(d), we compares the spin gaps of the $t$-$J$ and Hubbard models at 1/3 filling ($n=2/3$). 
The spin gap is defined as $\Delta_{\rm S} = E_0(N_{\uparrow}+1,N_{\downarrow}-1) - E_0(N_{\uparrow},N_{\downarrow})$, where $E_0(N_{\uparrow},N_{\downarrow})$ is the lowest energy of $N_{\uparrow}$ up-spin and $N_{\downarrow}$ down-spin electrons. 
To estimate the values in the thermodynamic limit, we performed an extrapolation of $\Delta_{\rm S}$ as shown in Fig.~\ref{fig5}(c). 
The spin gap of the $t$-$J$ model increases with $t_{\perp}/t_{\parallel}$ [Fig.~\ref{fig5}(d)]. 
On the other hand, when $t_{\perp}/t_{\parallel}=2.5$, the spin gap of the Hubbard model is much smaller than that of the $t$-$J$ model. 
The spin gap of the Hubbard model at $t_{\perp}/t_{\parallel}=2.5$ is also comparable to that of the $t$-$J$ model at $t_{\perp}/t_{\parallel}=1.5$. 
In terms of the interchain hopping $t_{\perp}$, $U/t_{\parallel}=10$ with $t_{\perp}/t_{\parallel}=2.5$ correspond to $U/t_{\perp}=4$, which may be too small to fully enable the $t$-$J$ model description along the rung direction. 
The difference in $\Delta_{\rm S}$ at $t_{\perp}/t_{\parallel}=2.5$ is caused by the difference in the ability of double occupancy in the Hubbard and $t$-$J$ models. 
Although the three-site hopping term proportional to $J$ is omitted in the $t$-$J$ Hamiltonian of Eq.~\eqref{eq:H_t-J}, it lowers the energy of the three-site state of Eq.~\eqref{eq:three_site} and may not substantially bridge the gap between the Hubbard and $t$-$J$ models. 
If the spin gap, which is favorable for the pair formation, can be made sufficiently large in the Hubbard model, the pair correlations may develop similarly to those in the $t$-$J$ model.

\section{Summary} \label{sec:summary}

We investigated the pairing properties of the three-leg $t$-$J$ and Hubbard ladders near 1/3 filling using the DMRG method. 
In the hole-doped system, the pair correlation functions exhibit power-law decays, whereas the spin correlation functions decay exponentially. 
On the other hand, in the case of electron doping, the pair correlation functions do not develop substantially. 
Hence, the hole-doped state near 1/3 filling is suitable for the emergence of SC. 
This asymmetric pairing nature across the 1/3-filling line in the $t$-$J$ model differs from the phase diagram predicted in the weak-coupling limit~\cite{EArrigoni1996_1,EArrigoni1996_2,HLin1997}. 
For hole-doped states, we showed changes in the ground-state properties as a function of $t_{\perp}/t_{\parallel}$ and discussed the pairing properties in the three-leg Hubbard ladder. 

The doped state we studied in the three-leg ladder near 1/3 filling is similar to the pairing states in the two-leg ladders near half filling that involve interchain coupling~\cite{EDagotto1992,RNoack1996,DKato2020,ASheikhan2020}. 
Compared to the two-leg Hubbard and $t$-$J$ models, the presence of a hole in each unit cell [or three-site state of Eq.~\eqref{eq:three_site}] likely makes it more difficult for the pairing state to emerge near 1/3 filling in the three-leg model. 
In trilayer nickelates, the $d_{3z^2-r^2}$ orbitals are nearly 1/3 filled~\cite{HSakakibara2024_4310}. 
If the $d_{3z^2-r^2}$ orbitals are responsible for pair formation, our discussion using the three-leg ladder may partially capture the pairing mechanism in trilayer nickelates. 
Recent calculations using the fluctuation-exchange approximation in the trilayer Hubbard model have also revealed a similar tendency; the hole-doped region near 1/3 filling is more favorable for SC than electron doping~\cite{YYamada_flex}. 
For trilayer nickelates, it is necessary to consider the itinerant electrons in the $d_{x^2-y^2}$-orbital network; integrating other contributions present in trilayer nickelates will be an important research topic in the future.

\begin{acknowledgments}
We thank M.~Ochi for fruitful discussions. 
This work was supported by Grants-in-Aid for Scientific Research from JSPS, KAKENHI Grant No.~JP24K06939, No.~JP24H00191, No.~JP24K01333, and No.~JP25H01252. 
M.K. was supported by the JSPS Research Fellowship for Young Scientists and Grant-in-Aid for JSPS Fellows Grant No.~JP25KJ1758. 
R.U. was supported by the Program for Leading Graduate Schools: ``Interactive Materials Science Cadet Program'' and JST SPRING, Grant No.~JPMJSP2138. 
The DMRG calculations were performed using the ITensor library~\cite{ITensor,ITensor2}.
\end{acknowledgments}

\appendix 
\section{Three-site states} \label{appendix_A}

Here, we address the lowest-energy state of three sites in each unit cell, given by the Hamiltonian 
\begin{align}
\hat{H} = 
& -t_{\perp} \sum_{\langle l, l' \rangle} \sum_{\sigma} 
\left( \hat{\tilde{c}}^{\dagger}_{l,\sigma} \hat{\tilde{c}}_{l',\sigma} + \mathrm{H.c.} \right)
\notag \\
& +J_{\perp} \sum_{\langle l, l' \rangle} 
\left( \hat{\bm{S}}_{l}\cdot \hat{\bm{S}}_{l'} - \frac{1}{4}\hat{n}_{l}\hat{n}_{l'} \right). 
\end{align}
At 1/3 filling, the eigenstates with one up spin and one down spin can be configured by 
\begin{align}
\ket{1_{\pm}} = \frac{1}{\sqrt{2}} \left( \ket{\uparrow,\downarrow,0} \pm \ket{\downarrow,\uparrow,0} \right), 
\\
\ket{2_{\pm}} = \frac{1}{\sqrt{2}} \left( \ket{0,\uparrow,\downarrow} \pm \ket{0,\downarrow,\uparrow} \right), 
\\
\ket{3_{\pm}} = \frac{1}{\sqrt{2}} \left( \ket{\uparrow,0,\downarrow} \pm \ket{\downarrow,0,\uparrow} \right), 
\end{align}
where $\ket{\uparrow,\downarrow,0} = \hat{c}^{\dagger}_{1,\uparrow}\hat{c}^{\dagger}_{2,\downarrow}\ket{0}$, $\ket{0,\uparrow,\downarrow} = \hat{c}^{\dagger}_{2,\uparrow}\hat{c}^{\dagger}_{3,\downarrow}\ket{0}$, $\ket{\uparrow,0,\downarrow} = \hat{c}^{\dagger}_{1,\uparrow}\hat{c}^{\dagger}_{3,\downarrow}\ket{0}$, $\ket{\downarrow,\uparrow,0} = \hat{c}^{\dagger}_{1,\downarrow}\hat{c}^{\dagger}_{2,\uparrow}\ket{0}$, $\ket{0,\downarrow,\uparrow} = \hat{c}^{\dagger}_{2,\downarrow}\hat{c}^{\dagger}_{3,\uparrow}\ket{0}$, and $\ket{\downarrow,0,\uparrow} = \hat{c}^{\dagger}_{1,\downarrow}\hat{c}^{\dagger}_{3 \uparrow}\ket{0}$. 
$\ket{\xi_+}$ ($\xi=1,2,3$) corresponds to the spin-triplet ($S=1$) state with $S^z=0$, while $\ket{\xi_-}$ corresponds to the spin-singlet ($S=0$) state. 
Since $\braket{\xi_+| \hat{H} | \xi'_-}=0$, the eigenstates at 1/3 filling are given by two matrices 
\begin{equation}
H_+ = \left[
\begin{array}{ccc}
  0          & 0          & -t_{\perp} \\
  0          & 0          & -t_{\perp} \\
  -t_{\perp} & -t_{\perp} & 0          \\
\end{array}
\right], \;
H_- = \left[
\begin{array}{ccc}
  -J_{\perp} & 0          & -t_{\perp} \\
   0         & -J_{\perp} & -t_{\perp} \\
  -t_{\perp} & -t_{\perp} & 0
\end{array}
\right]. 
\end{equation}
When $J_{\perp} >0$, the matrix $H_-$ gives the lowest energy 
\begin{align}
E_{0}(N=2) = \frac{-J_{\perp}-\sqrt{J_{\perp}^2+8t_{\perp}^2}}{2}, 
\end{align}
whose eigenstate is 
\begin{align}
\ket{\psi_0} = \frac{u}{\sqrt{2}}\ket{1_-} + \frac{u}{\sqrt{2}}\ket{2_-} + v \ket{3_-}, 
\end{align}
where
\begin{equation}
u = \sqrt{ \frac{1}{2} \left(1 \!+\! \frac{J_{\perp}}{\sqrt{J_{\perp}^2 \!+\! 8t_{\perp}^2}} \right)}, 
\;
v = \sqrt{ \frac{1}{2} \left(1 \!-\! \frac{J_{\perp}}{\sqrt{J_{\perp}^2 \!+\! 8t_{\perp}^2}} \right)}. 
\end{equation}
The spin correlations of $\ket{\psi_0}$ are given by 
\begin{align}
&\braket{\hat{\bm{S}_1} \cdot \hat{\bm{S}}_2} 
= -\frac{3}{8}u^2 
= -\frac{3}{16} \left(1 \!+\! \frac{J_{\perp}}{\sqrt{J_{\perp}^2 \!+\! 8t_{\perp}^2}} \right) , 
\\
&\braket{\hat{\bm{S}_1} \cdot \hat{\bm{S}}_3} 
= -\frac{3}{4}v^2 
= -\frac{3}{8} \left(1 \!-\! \frac{J_{\perp}}{\sqrt{J_{\perp}^2 \!+\! 8t_{\perp}^2}} \right). 
\end{align}

\bibliography{reference}

\end{document}